\documentclass{article}  

\let\ssection=\section
\renewcommand{\section}{\setcounter{equation}{0}\ssection}
\usepackage{graphicx}
\usepackage{amsmath,amssymb,amsfonts}

\title{On the non-linearity of the subsidiary systems}
\author {Helmut Friedrich\\
Albert-Einstein-Institut\\
Max-Planck-Institut f\"ur Gravitationsphysik\\
Am M\"uhlenberg 1\\
14476 Golm\\ 
Germany }

\begin{document}           

\maketitle                 
 
\begin{abstract}
\noindent
In hyperbolic reductions of the Einstein equations the evolution of 
gauge conditions or constraint quantities is controlled by subsidiary systems.
We point out a class of non-linearities in these systems which may have the
potential of generating catastrophic growth of gauge resp. constraint violations in
numerical calculations.
\end{abstract}

\vspace{3cm}

\section{Introduction}

Most numerical calculations of solutions to Einstein's field equations are being
plagued by an undesirably fast growth of constraint violations. A huge variety
of reduced equations has been derived from Einstein's equations with the hope of
finding versions with stable propagation properties and there have
been suggested modifcations of the equations which were hoped to force back the
solution to the constraint manifold (we refer to \cite{lehner:reula}
for further dicussion and references). More recently, there have been performed
stability analyses of  the subsidiary systems which control the evolution of
the gauge conditions or the constraint quantities (cf. \cite{frauendiener:vogel},
\cite{yoneda:shinkai} and the references given there). These led to requirements
on the coefficients of the equations which in the case of
\cite{frauendiener:vogel} resulted in geometric conditions on the foliation
underlying the evolution by the main system. Nevertheless, the field appears to
be wide open, most of the suggested remedies are experimental, and the cause of the
problems is not understood. In
\cite{frauendiener:vogel}, \cite{yoneda:shinkai} the subsidiary systems have been
considered as linear systems on given space-times.  In contrast, we wish to
emphasize in this note the non-linearity of the subsidiary system, which appears
to have a potential of generating catastrophic constraint violations. To
understand the extent to which these non-linearities may affect the numerical
construction of space-times and to develop, if necessary, ways to avoid
their effects, further investigations are needed. 

\section{The hyperbolic reduction}

We shall use a few facts about the {\it hyperbolic reduction} procedure
by which the geometric initial value problem for Einstein's field equations is
reduced to a Cauchy problem for a hyperbolic system. There exist now many
versions of it, whose general underlying structure are, however, more or less the
same. Because it leads to concise expressions, we will use the metric
coeffficients as basic unknowns and consider the representation of the field
equations in which their evolution is governed by a system of wave equations. To
emphasize the independence of our discussion of any particular coordinate system
we shall employ the notion of a gauge source function introduced in
\cite{friedrich:1hyp red}. We refer to  \cite{friedrich:2hyp red},
\cite{friedrich:rendall} for more details on the reduction procedure.

Let $(M, g)$ denote a smooth $4$-dimensional Lorentz space with smooth
space-like Cauchy hypersurface $S$ and $U \ni x^{\lambda} \rightarrow
F^{\mu}(x^{\lambda}) \in V$ a smooth map of an open subset $U$ into another
subset $V$ of $\mathbb{R}^4$. With functions $x^{\mu}$ and their
(4-dimensional) differentials
$dx^{\nu}$ prescribed on some open subset $W$ of $S$ one can solve near $W$ the
Cauchy problem for the semi-linear system of wave equations
\[
\nabla_{\nu}\,\nabla^{\nu}\,x^{\mu} = - F^{\mu}(x^{\lambda}).
\]
If the $dx^{\mu}$ have been chosen linearly independent on $W$ the solution will
provide a smooth coordinate system $x^{\lambda}$ on some neighbourhood
of $W$ in $M$. In terms of these coordinates the equations above take the form
\[
- \Gamma^{\mu}(x^{\lambda}) = - F^{\mu}(x^{\lambda}),
\]
where the $\Gamma^{\mu}$ denote the contracted Christoffel symbols
$\Gamma^{\mu} = g^{\nu \eta}\,\Gamma_{\nu}\,^{\mu}\,_{\eta}$ of $g_{\mu\nu}$. This
shows (ignoring subtleties arising in situations of low differentiability) that the
contracted Christoffel symbols can locally be made to agree with any prescribed set
of functions $F^{\mu}$ and that these function and the initial data determine the
coordinates uniquely. We refer to these functions as {\it gauge source functions}.

Assume now that $(M, g)$ is to be obtained by solving a Cauchy problem
for Einstein's vacuum field equations. We shall derive the properties
which will help us formulate this problem as a Cauchy problem for hyperbolic
equations.  The contracted Christoffel symbols are of particular interest to us
because the Ricci tensor of $g$ can be written in the form
\begin{equation}
\label{einlanc}
R_{\mu \nu} = - \frac{1}{2}\,g^{\lambda \rho}\,g_{\mu \nu,\lambda \rho} +
\nabla_{(\mu}\Gamma_{\nu)}
+ \Gamma_{\lambda}\,^{\eta}\,_\mu\,g_{\eta \delta}
\,g^{\lambda \rho}\,\Gamma_{\rho}\,^{\delta}\,_\nu
+ 2\,\Gamma_{\delta}\,^{\lambda}\,_{\eta}\,
g^{\delta \rho}\,g_{\lambda(\mu}\,\Gamma_{\nu)}\,^{\eta}\,_{\rho}.
\end{equation}
Here the contracted Christoffel symbols (and the functions 
$F_{\nu}  = g_{\nu \mu}\,F^{\mu}$ considered in the following) are being
formally treated as if they defined a vector field (which, of course, they do
not).  Thus $\Gamma_{\nu} = g_{\nu \mu}\,\Gamma^{\mu}$ and $
\nabla_{\mu}\Gamma_{\nu} =
\partial_{\mu}\Gamma_{\nu} -
\Gamma_{\mu}\,^{\lambda}\,_{\nu}\,\Gamma_{\lambda}$.

The discussion above suggests replacing in (\ref{einlanc}) the functions
$\Gamma_{\nu}$ by freely chosen gauge source functions $F_{\nu}$ (so that the
resulting expression will depend in general on the coordinates $x^{\lambda}$ not
any longer only through the $g_{\mu\nu}$). With this replacement the vacuum field
equations take the form 
\begin{equation}
\label{2redequ}
0 = R^F_{\mu\nu} \equiv - \frac{1}{2}\,g^{\lambda \rho}\,g_{\mu \nu,\lambda \rho}
+
\nabla_{(\mu}\,F_{\nu)}
+ \Gamma_{\lambda}\,^{\eta}\,_\mu\,g_{\eta \delta}
\,g^{\lambda \rho}\,\Gamma_{\rho}\,^{\delta}\,_\nu
+ 2\,\Gamma_{\delta}\,^{\lambda}\,_{\eta}\,
g^{\delta \rho}\,g_{\lambda(\mu}\,\Gamma_{\nu)}\,^{\eta}\,_{\rho},
\end{equation}
of a system of wave equations for the $g_{\mu \nu}$. We refer to
(\ref{2redequ}) as the {\it main evolution system} or the {\it reduced
equations}. For this system the Cauchy problem for $g_{\mu \nu}$ with data
satisfying the constraints on a space-like hypersurface
$S$ is well posed.

Suppose that $g_{\mu \nu}$ is a solution of this problem near $S$. Since
equation (\ref{2redequ}) is in fact of the form 
\begin{equation}
\label{3redequ}
R_{\mu \nu} = \nabla_{(\mu}\,Q_{\nu)},
\end{equation}
where $Q_{\mu} = \Gamma_{\mu} - F_{\mu}$ with the $\Gamma_{\mu}$ calculated 
from $g_{\mu \nu}$, it is not clear a priori whether the
solution $g$ will indeed satisfy the {\it gauge condition} $\Gamma_{\mu} =
F_{\mu}$ and thus the vacuum field equation
$R_{\mu \nu} = 0$.

The Bianchi identity, which holds for any metric, implies 
\begin{equation}
\label{subs}
0 = 2\,\nabla^{\mu}(R_{\mu \nu} - \frac{1}{2}\,R\,g_{\mu \nu})   
= \nabla_{\mu}\,\nabla^{\mu}\,Q_{\nu} + R^{\mu}\,_{\nu}\,Q_{\mu},
\end{equation}
and thus a system of wave equations for the quantities $Q_{\nu}$. We refer to
this system as {\it subsidiary system}. 

The detailed analysis shows that if the Cauchy problem for the main evolution
system is arranged so that the initial data satisfy the constraints and the
gauge condition $Q_{\nu} = 0$ on $S$, it follows for the solution of
the main evolution system that also $dQ_{\nu}$ and thus any `time derivative'
$\partial_t\,Q_{\nu}$ transverse to $S$ vanishes on $S$. The uniqueness
property of the subsidiary system  therefore implies that the solution to
the main evolution system does indeed satisfy $Q_{\nu} = 0$ on the
domain of dependence of $S$ with respect to $g$. This reduces the local
Cauchy problem for Einstein's field equations to the problem of solving equation
(\ref{2redequ}), which thus takes the central role in the analytic discussion.
Of the subsidiary system only the homogeneity and the resulting uniqueness
property are needed.

\section{The non-linearity of the subsidiary equation}

The  main evolution system is also central in numerical discussions.
Because the initial data $Q_{\mu}$ and $\partial_t\,Q_{\nu}$ on the
initial hypersurface $S$ come with an error, the evolution properties of the
subsidiary system will, however, also become important.

There is no way to relate a numerical solution to the solution of
the continuum problem one wants to approximate. To get some idea how errors
infiltrate into the various systems, it is useful to consider an analogy
accessible to analytic methods.  We assume the main evolution system to be
satisfied by fields $g_{\mu
\nu}$ of class $C^3$ with an error term $E_{\mu \nu}$ of class $C^1$ so
that 
\begin{equation}
\label{ERRredequ}
R_{\mu\nu} = \nabla_{(\mu}\,Q_{\nu)} + E_{\mu \nu}.
\end{equation}
Nothing will be assumed about the origin and structure of this error and the
errors in the initial data $Q_{\mu}$ and $\partial_t\,Q_{\nu}$ on $S$. 

Using the Bianchi identity with (\ref{ERRredequ}) gives the analogue
\begin{equation}
\label{INHOMsubs}
\nabla_{\mu}\,\nabla^{\mu}\,Q_{\nu} + R^{\mu}\,_{\nu}\,Q_{\mu} = 
- 2\,\nabla^{\mu}(E_{\mu \nu} - \frac{1}{2}\,g_{\mu
\nu}\,E_{\rho}\,^{\rho}),
\end{equation}
of the subsidiary equation (\ref{subs}). This equation will in general not be
homogeneous any longer.  There may be a way to avoid this problem.
Assume a   splitting of the form 
\[
E_{\mu \nu} = \nabla_{(\mu}\,e_{\nu)} + f_{\mu \nu},
\]
with some $C^2$ vector field $e_{\mu}$ and some symmetric $C^1$ tensor
field $f_{\mu \nu}$. With (\ref{ERRredequ}) this gives
\begin{equation}
\label{einstQ'f}
R_{\mu\nu} = \nabla_{(\mu}\,Q'_{\nu)} + f_{\mu \nu}
\,\,\,\,\,\,\,\mbox{with}\,\,\,\,\,\,\,
Q'_{\nu} = Q_{\nu} + e_{\nu}.
\end{equation}
If $f_{\mu \nu}$ would vanish this would just amount to a redefinition
\[
F_{\mu} \rightarrow F'_{\mu} = F_{\mu} - e_{\mu},
\] 
of the gauge source function which may be quite harmless. Moreover,
we would get a homogenous wave equation for $Q'_{\mu}$. If $E_{\mu\nu}$ is
known but not of the form $\nabla_{(\mu}\,e_{\nu)}$, a homogenous
system can be obtained by choosing the splitting suitably. 
If we require $e_{\mu}$ to solve the system of wave equations
\[
\nabla_{\mu}\,\nabla^{\mu}\,e_{\nu} + R^{\mu}\,_{\nu}\,e_{\mu} =
2\,\nabla^{\mu}\left(E_{\mu \nu} - \frac{1}{2}\,g_{\mu
\nu}\,E_{\rho}\,^{\rho}\right),
\]
it follows from the splitting above that  
\[
\nabla^{\mu}\left(f_{\mu \nu} - \frac{1}{2}\,g_{\mu
\nu}\,f_{\rho}\,^{\rho}\right) = 0,
\]
and the Bianchi identity implies for $Q'_{\mu}$ the homogeneous equation
\begin{equation}
\label{HOMsubs'}
\nabla_{\mu}\,\nabla^{\mu}\,Q'_{\nu} + R^{\mu}\,_{\nu}\,Q'_{\mu} = 0.
\end{equation}

One could now try to analyse how errors in $Q_{\mu}$ and
$\partial_t\,Q_{\nu}$ on $S$ are propagated, in dependence of $E_{\mu
\nu}$, by (\ref{INHOMsubs}) or how errors in $Q'_{\mu}$ and
$\partial_t\,Q'_{\nu}$ on $S$ will be propagated by (\ref{HOMsubs'}). 
Assuming in this analysis $E_{\mu\nu}$ and $g_{\mu\nu}$ whence $R_{\mu\nu}[g]$
as suitably bounded but given with no further information, standard
energy estimates admit but cannot exclude exponential growth of $Q_{\mu}$ resp.
$Q'_{\mu}$. One might be able to work out conditions on $R_{\mu\nu}[g]$ which will
allow one to keep the errors under control. However, by itself this will be 
of limited use. 
Inserting (\ref{ERRredequ}) into (\ref{INHOMsubs}) and, observing the
splitting, into (\ref{HOMsubs'}), gives
\begin{equation}
\label{nonlinINHOMsubs}
\nabla_{\mu}\,\nabla^{\mu}\,Q_{\nu} 
+ Q^{\lambda}(\nabla_{(\lambda}\,Q_{\nu)} + E_{\lambda \nu})
=  - 2\,\nabla^{\mu}(E_{\mu \nu} - \frac{1}{2}\,g_{\mu
\nu}\,E_{\rho}\,^{\rho}),
\end{equation}
and 
\begin{equation}
\label{nonlinHomsubs}
\nabla_{\mu}\,\nabla^{\mu}\,Q'_{\nu} + 
Q^{'\lambda}(\nabla_{(\lambda}\,Q'_{\nu)}
+ f_{\lambda \nu}) = 0,
\end{equation}
respectively. The relative size of the errors $E_{\lambda \nu}$ and
$f_{\lambda \nu}$ is not clear. More important is that in this form the 
subsidiary system shows its non-linearity. 

If $Q'_{\mu}$ and $\partial_t\,Q'_{\nu}$ vanish on $S$, equation
(\ref{nonlinHomsubs}) will still imply that $Q'_{\mu}$ vanishes in the domain
of dependence. 
If $Q'_{\mu}$ and $\partial_t\,Q'_{\nu}$ do not vanish on $S$, however, the
following observations indicate that (\ref{nonlinHomsubs}) may imply a growth of the
solution $Q'_{\mu}$ which is worth than exponential. We assume
$g = dt^2 - \delta_{ab}\,dx^a\,dx^b$ (so that (\ref{nonlinHomsubs}) 
decouples from (\ref{einstQ'f})), $f_{\lambda \nu} = 0$,
and initial data which satisfy for $a = 1, 2, 3,$
\[
Q'_{a} = 0, \,\,\,\, \partial_t Q'_{a} = 0, \,\,\,\,
\partial_a\,Q'_0 = 0,\,\,\,\,
\partial_a\,\partial_tQ'_0 = 0
\quad \mbox{on}\quad \{t = 0\},
\]
so that the error resides only in the constant functions 
$a = Q'_0$, $b = \partial_tQ'_0$ on $\{t = 0\}$. 

Equation (\ref{nonlinHomsubs}) then implies  $Q'_a \equiv 0$ and reduces in
fact to
$\partial_t\,Q'_{0} = \frac{1}{2}\,(c - Q_{0}^{'2})$ with $c = 2\,b + a^2$.
The integration gives 
$Q_0' = a = const.$ if $b = 0$, 
$Q'_0 = \sqrt{c}\,\,\frac{a + \sqrt{c}\,\tanh 
\frac{\sqrt{c}\,t}{2}}{\sqrt{c} + a\,\tanh \frac{\sqrt{c}\,t}{2}}$ if 
$0 \neq 2\,b > - a^2$,
$Q'_0 = \frac{2\,a}{a\,t + 2}$ if $a^2 = - 2\,b$, and
$Q'_0 = \sqrt{|c|}\,\tan \left\{- \frac{\sqrt{|c|}\,t}{2}
+ \arctan \frac{a}{\sqrt{|c|}} \right\}$ if $2\,b < - a^2$.
The solutions thus remain bounded for $t \ge 0$ if $b \ge 0$ or if $a \ge
0$ and $- a^2 < 2\,b < 0$, while they develop {\it poles} at some $t_* >
0$ if $b < 0$ and $a \le 0$ or if $2\,b < - a^2$ and $a > 0$.
(If the corresponding initial data would be modified outside the intersection of the
hypersurface $\{t = 0\}$ with the backward light cone of the point
$(t_*, x^a)$, the solution would still become singular at $(t_*, x^a)$.)

We note that in general any Killing vector field $K$ of a vacuum
solution $g_{\mu \nu}$ satisfies equation
(\ref{subs}) with $Q^{\mu} = K^{\mu}$ in the coordinates $x^{\mu}$. In the
present case this gives solutions which grow linearly. 

\vspace{.2cm}

For us the following observation is important:

\vspace{.1cm}

{\it With our assumption on $g_{\mu\nu}$ it follows that each neighbourhood of
the initial data 
$Q_{a} = 0$ and $\partial_t Q_{a} = 0$ contains
initial data for (\ref{subs}) for which the solution $Q_{a}(x^{\mu})$ become
unbounded at some finite $x^0 = t_* > 0$}.

\vspace{.3cm}

\newpage

\section{Concluding remarks}

The only purpose of the discussion above was to indicate the behaviour of
solutions to the subsidiary system. We expect to find a similar singular
behaviour of the solutions to equation (\ref{subs}) if the latter is given with a
general smooth metric $g_{\mu\nu}$ (extending sufficiently far into the future).
It would be interesting to know whether the {\it singular data set}, i.e. the subset
of data which determine singular solutions of equation (\ref{subs}), is open or of
lower dimension in the set of all data if the metric $g$ is given. While the
singular data set itself will depend on the metric $g$, such a characterization may
be independent of the chosen metric and may help avoid entering the singular
sector. 

In a consistent discussion of the growth of $Q_{\mu}$ one would have to consider
the main system and the subsidiary system
\[
R_{\mu \nu} = \nabla_{(\mu}\,Q_{\nu)}, \quad\quad
\nabla_{\mu}\,\nabla^{\mu}\,Q_{\nu}  +
Q^{\lambda}\nabla_{(\lambda}\,Q_{\nu)} = 0,
\]
as a coupled system. The main equation can be studied independently. The
subsidiary system, implicit in the main system, depends on the metric defined by
the latter. If $Q_{\mu}$ tends to grow, the main system will react to
it, whether for the better or the worth is not clear. 
Whether with changing $g_{\mu \nu}$ the non-linearity of the
subsidiary system can still imply a blow up of $Q_{\mu}$ at a finite time
needs to be analysed. 

For this purpose it might be interesting to study under simplifying
assumptions such as spherical symmetry whether the model system
\[
R_{\mu \nu} = \nabla_{(\mu}\,q_{\nu)}, \quad\quad
\nabla_{\mu}\,\nabla^{\mu}\,q_{\nu}  +
q^{\lambda}\nabla_{(\lambda}\,q_{\nu)} = 0,
\]
considered as Einstein equations coupled to a source
field given by a {\it vector field} $q_{\mu}$, will develop
a blow up for suitable data.  The reduction of these equations is obtained by a
slight modification of the one described above. The system can be
simplified further by assuming $q_{\mu}$ to be a differential
$q_{\mu} =
\nabla_{\mu}\,f$ of some function $f$. The second equation will then be
implied if $f$ satisfies
\[
\nabla_{\mu}\,\nabla^{\mu}f  + \nabla_{\mu}f\,\nabla^{\mu}f = const.,
\]
which is a wave map equation in the case where the constant on the right
hand side vanishes. In any case the results might lead to an identifcation of a
mechanism responsible for the growth of constraint violations and to the
development of methods to avoid them.

Some authors add terms built from $Q_{\nu}$ to the main system, which
appear to reduce the growth of the constraint violations in certain calculation
\cite{gundlach}, \cite{pretorius}. This may be related to the different effects of
the non-linearities which are obtained in the appropriately modified
subsidiary systems.  

There are available now many different types of reductions. Depending
on several choices, the subsidiary system may control the preservation of the
gauge or the preservation of constraints or a mixture thereof. In spite of the
different appearance of the resulting  main and subsidiary systems we expect
that similar non-linearities as the one discussed above will occur in any
subsidiary system, though, depending on the system, they may have different
effects.

In a numerical scheme for the second order wave equations 
the subsidiary system, which is of third order in the metric, can, of course, 
hardly be identified any longer as a kind of identity and the relations between
the two systems is obscured. But if the non-linearity of the subsidiary system
can have for non-vanishing initial data $Q_{\mu}$ and $\partial_t\,Q_{\mu}$
drastic effects in the continuum model, they are likely to be reflected 
in numerical calculations.

\end{document}